# Federated Learning: Attacks, Defenses, Opportunities, and Challenges

Ghazaleh Shirvani[1], Saeid Ghasemshirazi[2] and Behzad Beigzadeh[3]
[1] School of Computer Science, Carleton University, Ottawa, Canada
[2] School of Computer Science, Carleton University, Ottawa, Canada
[3] Department of Electrical and Computer Engineering, Tarbiat Modares University, Tehran, Iran

E-mail: ghazalehshirvani@cmail.carleton.ca, saeidghasemshirazi@cmail.carleton.ca, behzadbeigz@yahoo.com

*Abstract*— **Using dispersed data and training, federated learning (FL) moves AI capabilities to edge devices or does tasks locally. Many consider FL the start of a new era in AI, yet it is still immature. FL has not garnered the community's trust since its security, and privacy implications are controversial FL's security and privacy concerns must be discovered, analyzed, and recorded before widespread usage and adoption. A solid comprehension of risk variables allows an FL practitioner to construct a secure environment and provide researchers with a clear perspective of potential study fields, making FL the best solution in situations where security and privacy are primary issues. This research aims to deliver a complete overview of FL's security and privacy features to help bridge the gap between current federated AI and broad adoption in the future. In this paper, we present a comprehensive overview of the attack surface to investigate FL's existing challenges and defense measures to evaluate its robustness and reliability. According to our study, security concerns regarding FL are more frequent than privacy issues. Communication bottlenecks, poisoning, and backdoor attacks represent FL's privacy's most significant security threats. In the final part, we detail future research that will assist FL in adapting to real-world settings.**

## I. INTRODUCTION

THANKS to the expansion of IoT devices [1], cell phones, and huge websites throughout the world, tremendous amounts of data have been produced and shared in recent years. Moving raw data from individual devices or the data centers of different firms to a centralized server or data center may present an immediate or potential information leakage risk if the data contain sensitive information about end users or organizations, such as location-based services, health information, or personal financial standing. Data security and privacy concerns have prompted the development of several regulatory restrictions[2], also include the US Consumer Privacy Bill of Rights (CPBR)[3].
Furthermore, computer and storage resources are often dispersed across different locations and companies[4], implying that they cannot be collected centrally in a single data center.

The emergence of Federated Learning as a viable method for cooperatively training a machine learning model using dispersed resources is a significant step forward in this direction. To train a model, FL uses a distributed machine learning technique in which numerous users work together without moving the unprocessed data to a centralized server or data center[5,6]. As a bonus, FL asserts that it guarantees the privacy and security of decentralized raw data storage.

In FL, participants are interested in learning about a global model. Amazon Alexa users, for instance, have a vested interest in using the most cutting-edge voice recognition technology feasible on their gadgets[7]. Another well-appropriate instance would be Intelligent Healthcare programs. The primary target of this perspective is learning vital metrics of prevalent patience with similar diseases and utilizing them for a novel treatment far beyond existing technology and human processing power. In the past, users had to submit their local data to a centralized server to participate in machine learning training. Then, a machine learning model is trained on the centralized server.

Nonetheless, transferring the original data is usually not an option. In fact, users' reluctance to transmit data is mainly motivated by their desire to protect their personal information and is compounded by the presence of legal hurdles. Also, communication costs increase when delivering more significant amounts of data, such as multimedia files. Finally, it could be impossible to store and train on all the received data owing to computational limitations when there may be millions of participants[8].

Learning in a federated approach provides several benefits to overcome these issues. The key concept behind federated learning is that users would build a model locally on their own devices and then share the trained model rather than the data used to train it. After receiving models from users, a central server may combine them into a single global model before sending it back. The received model initiates a fresh training cycle for the user. In this way, Models (which consist of results from the locally trained user's data) are trained via iterations of this process. So that data would be kept secret, and most calculations could be performed in a distributed manner. However, several more opportunities exist for an adversary to launch attacks against this system. An adversary may now easily affect a model's training process. An adversary may achieve many different goals by providing possibly arbitrary modifications to the primary server[9].

Firstly, there is a high probability of an attack since there are many potential targets with federated learning devices (millions of individuals). Due to this, we conclude that attacks against federated learning systems are very probable. Indeed, it is essential to remember that the potential risk caused by attacks on federated learning differs from application to application. For example, the danger is minimal for next-word predictions in smart keyboards since, at worst, the user will be somewhat frustrated. However, the danger may be far more substantial in other applications like malware detection[10], healthcare, or smart grids.

Federate learning is susceptible to hostile attacks due to a lack of study into such attacks. Given the widespread of federated learning, we believe that it is crucial to have a deeper understanding of its security issues and to apply more strict mechanisms to enhance the security of the federated learning system. In particular, we focus on federated learning situations in which adversarial training is used. One of the most influential and versatile defense methods against adversarial instances is adversarial training. The objective of adversarial training, however, is more complicated than that of traditional training. In addition, secure aggregation approaches that protect against attackers in federated learning include a stringent screening of the provided updates[11]. In light of this, the feasibility of training with antagonistic clients remains a question to which we expect a definitive response will be provided by further investigation.

This study comprehensively surveys the Taxonomy of federated learning security, attack mechanisms, and defense techniques. The main contributions of this paper are:

The paucity of surveys that systematically classify and analyze security and privacy concerns is notable in the area of Federated Learning[12]. The vast majority of research undertaken in this sector merely supported different gradient values to enhance efficiency or proposed a more trustworthy model of FL. However, it is difficult to protect against all risks if not informed of all existent attacks and defensive categories and kinds. With this gap, specific forks attract more attention while others fade from view. We undertook this extensive survey in Federated Learning security to address this issue and provide valuable direction for future study in this field.

### A. Federated Learning

The proliferation of the need to process data has necessitated distributed machine learning methods. Although distributed ML tackled the resource limitation problem, data privacy principles were tentative. Therefore, Federated Learning was introduced as a novel distributed machine learning technique involving numerous nodes working together to create a unified machine learning model. The breakthrough occurred when these nodes were linked without sharing information[12,13]. In other words, local datasets would be used to train local models, and then the critical parameters of the trained models would be forwarded to a global model for updating.

However, even a brilliant decentralized paradigm like federated learning is only amusing in theory. A hostile opponent may take advantage of this dispersion by authenticating itself as a benign node and injecting malicious data, which can provoke applications to provide incorrect results. Additionally, the adversary may provide a malicious command, which is especially effective in distributed denial of service (DDoS) assaults like Sybil[14].

TABLE 1
A LIST OF ABBREVIATIONS USED IN THIS SURVEY

| | |
|------|------------------------------------------|
| AI   | Artificial Intelligence                  |
| ML   | Machine Learning                         |
| FL   | Federated Learning                       |
| IID  | Independent and Identically Distributed  |
| IoT  | Internet of Things                       |
| HFL  | Horizontally Federated Learning          |
| BFL  | Blockchain-based FL                      |
| VFL  | Vertically Federated Learning            |
| FTL  | Federated Transfer Learning              |
| GAN  | Generative Adversarial Network           |
| IoV  | Internet of Vehicles                     |
| BFL  | Blockchain-based Federated Learning      |
| SMPC | Secure Multiparty Computation            |
| SGD  | Stochastic Gradient Descent              |
| NLP  | Natural Language Processing              |
| EC   | Edge Computing                           |
| HE   | Homomorphic Encryption                   |
| SMC  | Secure Multi-party computation           |
| DP   | Differential Privacy                     |
| SVM  | Support vector machine                   |
| LLR  | Log-Likelihood Ratio                     |

## II. BACKGROUND

### A. Federated learning implementation

The Federated Learning paradigm can be considered a repetitive learning process in which the global model is refined with each cycle[15]. The FL process flow comprises three stages:

- **Model initialization:** The initial ML model is delivered to each client's device via the FL server.
- **Local model training:** Each client device trains its own model using client-specific training data.
- **Aggregation of local models:** The FL server gathers updated model weights and combines them into the global model, which is then used to replace each client's local model.

As depicted in Fig 1, FL is engaged in a continuously progressive learning process that replicates prior learning stages (steps 2 and 3) to keep the global model current for all participants.

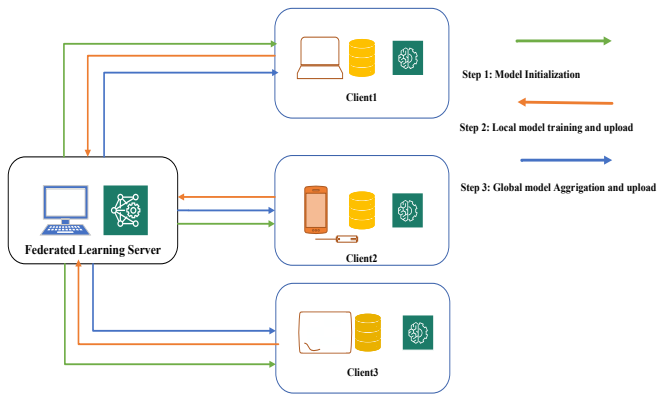

**Fig. 1.** A visual representation of the Federated Learning process, which demonstrates the decentralized approach of multiple parties collaboratively training a machine learning model.

For the deployment of FL, several prevalent datasets will come in handy. Table 2 is an overview of the publicly accessible datasets, some of which were accessible to the public and some not.

TABLE 2
AVAILABLE DATASETS FOR FEDERATED LEARNING DEPLOYMENT

| Dataset Name | Number of Items | Number of Clients | Reference |
|---|---|---|---|
| Shakespeare | 16068 | 715 | [16] |
| Feminist | 805,263 | 3550 | [17] |
| Federated EMNIST | 671,585 | 3400 | [18] |
| Celeba | 200288 | 9343 | [19] |
| CIFAR-100 | 60000 | 500 | [20] |

### B. TYPES OF FEDERATED LEARNING

#### 1) Horizontal Federated Learning

One of the most researched categories of federated learning is Horizontal Federated Learning (HFL). HFL is utilized when each device includes a dataset with the same feature space but distinct sample examples. On the downside, HFL can only put into practice applications of companies that compete over identical interests. Nevertheless, it is suitable for applications that apply to enormous heterogeneous devices with complex distributed networks. For instance, the initial use case of the FL-Google keyboard utilizes this form of learning in which mobile phones with identical characteristics but different training data participate[21,22].

#### 2) Vertical Federated Learning

On the contrary, Vertical Federated Learning (VFL) is applied when each device holds a dataset with individual properties but from the same sample instances. It means two organizations with data on the same group of people but different feature sets may build a shared ML model using Vertical FL. On the bright side, Vertical Federated Learning does not disclose people whose profiles do not overlap among participating parties[23].

In VFL, an alignment based on encryption is utilized to distinguish between two or more individuals with the same IDD. Then it has to gather this information since different parties have access to various aspects of the same individual. Simultaneously, VFL needs to maintain the user's anonymity; therefore, it has to compute the training loss and gradients. To sum up, in comparison to HFL, VFL is able to exploit more attribute dimensions. Hence, it calls for fewer resources which makes VFL more scalable. Despite all the qualities mentioned above of VFL, research in this branch of Federated Learning has been sparse[24].

#### 3) Federated Transfer Learning

Federated Transfer Learning (FTL) is another subset of federated learning distinct from horizontal and vertical FL. In FTL, the feature space of the two datasets differs. FTL applies to datasets obtained from distinct but comparable organizations. Given the distinctions in business nature, there is a slight overlap in feature space between these organizations. Furthermore, FTL applies to businesses established on a global scale. In such cases, datasets differ in samples and feature space[24].

In other words, Federated Transfer Learning resembles traditional Machine Learning in that a new feature is added to an already-trained model. Utilizing generic datasets as a starting point and then training the model on the sparse data to tailor it to a particular task. Prior to engaging in transfer learning, it is necessary to identify regions of resemblance between the source and target domains[25]. The purpose of transfer learning is to construct effective application-specific models in circumstances when data is insufficient. This is done using previously trained and proven successful models in a source domain closely comparable to the target domain. Then, with the help of the available data, the model may be oriented for usage in the intended application area. The different categories of FL have represented in Fig 2.

#### 4) Healthcare

Healthcare concerns would be federated learning's primary application. We need distributed algorithms like FL because of the enormous amount of essential data for disease detection or treatment training models. Since health information is very susceptible and its use is rigorously restricted, it is practically impossible to reach out to this data. Even if data anonymization could circumvent these restrictions, it is already common knowledge that deleting metadata such as the patient's name or date of birth is frequently insufficient to protect privacy[8,26,27]. Correspondingly collecting, organizing, and maintaining a high-quality data set is time-consuming, labor-intensive, and expensive; This is another reason why data sharing is not routine in the healthcare industry[28].

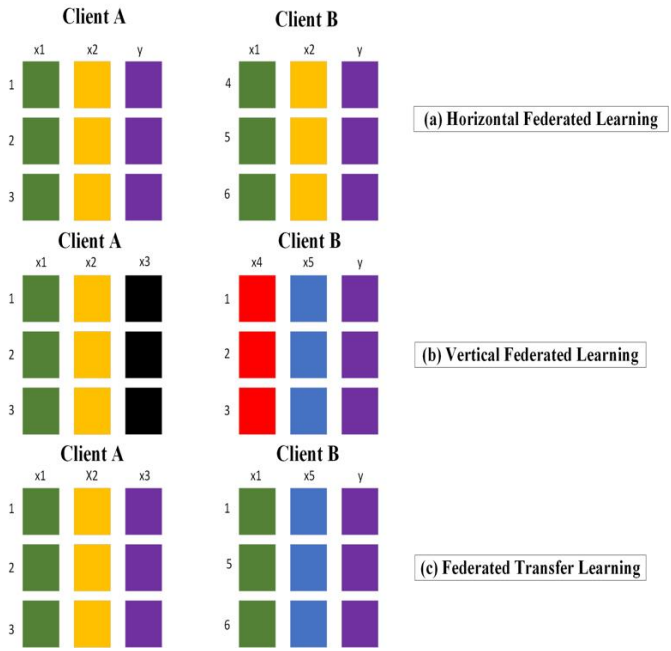

**Fig. 2.** Categorization of Federated Learning, including horizontal, vertical, and federated transfer learning. The figure's notation indicates that $x_n$ represents the $n_{th}$ feature of a sample's feature descriptions, y represents the sample label, and n represents the sample identifier[29].

### C. Federated Learning Application

#### 1) Insurance Section

Insurance companies collect and maintain a massive volume of data on a wide variety of subjects, including but not limited to policyholders' medical histories, vehicle and mobile phone details, corporate assets, and more. In addition, their insured may work with various companies[27]. Consequently, the challenge arises of training Machine Learning algorithms using distinct data sets when they cannot be shared across enterprises or locations. This problem is one that Federated Learning targets to solve. By utilizing federated learning, a business could discover the patterns of its users without breaking the data clause. Furthermore, by instituting Federal Learning, fraudulent or illegal behavior will be prevented. Therefore, without compromising the privacy of the insured, the data might be utilized to train and regulate the algorithms[3,30].

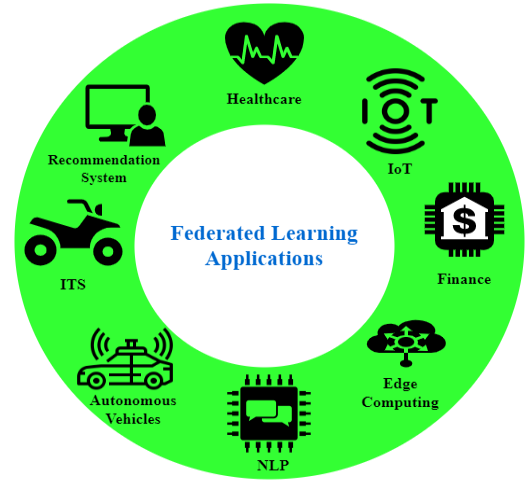

**Fig. 3.** Presentation of Top Federated Learning Applications

#### 2) Intelligent Transportation Systems

The second most probable application of federated learning can belong to IoV, especially Intelligent Transportation. FL has excellent potential to make this manufacturing more practical and secure. In this regard, Federated Learning is a game changer for next-generation networking technologies due to the wide variety of use cases it supports, such include Roadside Unit Intelligence, Network Function Virtualization Management and Orchestration, and Vehicular Intelligence[31].

#### 3) Banking and Finance

It is feasible to employ federated learning in finance. The value of collected information could be significantly enhanced by sharing it across other organizations. Customized and high-value services, such as automated stock or currency market tracking and trading, are powerful tools for financial institutions looking to boost client loyalty and pleasure. Credit scoring, risk management, and other business issues like Fraud can all be automated using AI models. However, most of this data includes a substantial amount of sensitive information, and the unpredictable impacts of data leakage complicate the benefits of information sharing and exchange[32]. Therefore, we require applying approaches similar to FL to introduce a secure way of training AI models; in this way, banks may collaborate without disclosing any sensitive information[33].

#### 4) Natural Language Processing

A Language Model (known as LM) is a model that estimates the probability of word sequences using an unstructured distribution. As a fundamental component of NLP systems, LM has been utilized in various NLP activities, including machine translation, text classification, correlation extraction, and knowledge discovery. Recent research on language modeling in Federated NLP focuses mainly on tackling a word-level LM challenge in the mobile market[34,35]. The majority of LMs in federated learning are implemented on the Google Keyboard, a virtual mobile keyboard. An example of a mobile keyboard suggestion is that models strive for mobile keyboard recommendations to be more dependable and resilient. Moreover, current models give quality increases in typing or expression (emoji) areas, such as future predictions, emoji predictions, query recommendations, etc.[34].

#### 5) Internet Of Things

Recently, the concept of FL for developing IoT systems with superior intelligence and privacy has been established.

By coordinating several devices with a central server, FL enables data training. In intelligent IoT networks, devices communicate with an aggregator (e.g., a server) to train neural networks as workers[36].

In particular, the aggregator launches a global model with learning parameters. Each worker retrieves the current model from the aggregator and then updates its model using its local dataset (e.g., stochastic gradient descent (SGD)); then offloads the local update to the aggregator. The aggregator then integrates all local model updates and builds a brand-new, enhanced global model. By leveraging dispersed workers' processing capacity, the aggregator can improve the quality of training while limiting user privacy leaks. Eventually, local employees receive the global update from the aggregator and calculate their local update until global training is complete[37,38].

#### 6) Autonomous Vehicles

Understanding and enhancing the working mechanism of blockchain and the federal approach is becoming crucial in vehicular networks because of the predicted growth of linked autonomous vehicles with varying service requirements. Federated learning reduces data transfer volume and expedites autonomous vehicle learning processes[39]. The advantages of networked autonomous vehicles include seamless message transmission and low-delay internet connections while driving[31,40]. The intended result is a systematic strategy for such a design that can convert our electric drivetrains into mobile data centers, execute FL, and rapidly respond to their needs.

#### 7) Recommendation System

Recommendation systems are heavily data-driven. Generally, the quality of suggestions generated by recommender systems improves as more data is used[31,40]. Due to privacy and security concerns, it is undesirable to share user information directly. Problems with silos and decentralization are likely to occur in recommender systems. Accordingly, numerous pilot studies have been conducted on maintaining data privacy and security when leveraging data silos. However, the majority of actions still demand that user data leave the local data storage. Federated learning is a novel technique that aims to combine data silos and construct machine learning models without jeopardizing user privacy or data security[22,38].

#### 8) Edge Computing:

FL is more convenient and less energy-consuming whenever the models are trained on edge devices. This statement implies that FL thrives in edge computing environments. This method makes training machine learning models on mobile edge networks easier. As a result, using FL inside the EC paradigm could help mitigate the costs of communication as well as the risks associated with privacy, legality, and security[36].

Additionally, FL enhances data collection while decreasing the associated costs and risks of centralized ML. Moreover, FL provides edge computing, considering it facilitates the group-based training of DL models for optimal network performance. There are advantages to using FL approaches on edge networks versus the more traditional centralized ML architecture. Data owners send update parameters to the FL server rather than raw data, significantly decreasing the volume and quantity of communication data[41].

Over this, it improves network bandwidth consumption. In time-sensitive applications (such as industrial control, mobility automation, navigation system, and real-time media), latency may be a major concern. Consequently, Local execution on end devices might also enhance the performance of real-time applications such as object tracking, virtual reality, and medical applications[42].

#### D. Threat Model

An essential step to comprehending attacks on FL methodology is ideally acknowledging the threat model. Threat models are structured representations of data that assist in detecting and describing possible security risks. They may be characterized in terms of the accessible data and the attacker's activity scope. Accordingly, we establish mutually exclusive concepts to describe the FL threat model.

#### 1) Security vs. Privacy.

When researching FL attacks, the most critical factor is whether they pose a system security risk or privacy. Depending on the application context, the repercussions of privacy attacks may weigh more than security attacks or vice versa. Data leakage, the core cause of privacy attacks, puts sensitive information at risk of falling into the wrong hands, whereas flaws in the system are the outcome of security attacks. For example, in the case of Intelligent Health, privacy attacks lead to patient data exposure, but in Intelligent vehicles, security attacks create inaccurate commands bringing financial and personal damages.

#### 2) Server vs. Client.

The primary distinction is the quantity of data at the disposal. Throughout each learning cycle, the server saves information about the model's architecture and client modifications. While client-based attacks only provide information about one or more clients, server-based attacks reveal more sensitive data.

#### 3) Adversary Knowledge.

Having white-box knowledge in a centralized setup is privy to all aspects of the target model, such as its parameters, architecture, and internal state. On the contrary, black-box knowledge means having almost zilch information about the target model. Also, there is a third type of adversary knowledge called grey-box knowledge. The grey-box attacker, first presented in [43], is a black-box attacker with some non-public statistical information about the target. Any node in an FL system, client or server alike, may be compromised by white-box, grey-box, or black-box attacks. Additionally, most attacks target client-owned data or communications between the federated server and clients. Therefore, we must ensure that we have all the accessible knowledge about the federated training process and client-owned data.

#### 4) Insider vs. Outsider.

The most general perspective in attack models is if the attack occurred from inside or outside[44]. Insider attacks are pretty harmful since they make adversaries able to modify valuable parameters or even the model's behavior. On the contrary, outsider attacks are about sniffing data (e.g., MITM or side-channel attacks). Insider attacks highlight two categories: Sybil and Byzantine.

● Sybil Attack:

Because of the consequences of distributed design, federated learning is vulnerable to saturation attacks, also known as Sybil attacks, from users. In Sybil attacks, the adversary increases the attack's success rate by managing several users. The users under Sybil's control will share comparable traits. The adversary influences the global model's training by manipulating a large number of users. Whereas the local model updates of reliable users frequently exhibit a variety of characteristics, local models operated by Sybils frequently exhibit the same characteristics[45]. Therefore, systems are susceptible to manipulation when servers lack detection capabilities.

● Byzantine Attack:

Byzantine attacks, one of the most prevalent attacks in federated learning, deliver arbitrary updates to the server that hinder the global learning mode's efficiency. They attempt to prevent the convergence of FL models by uploading random updates[46].

### III. FEDERATED LEARNING ADVERSARIAL ATTACK SURFACE TAXONOMY

One of the most severe attacks on federal models is recognized as adversarial attacks. Adversarial attacks can jeopardize almost all aspects of data security in models. Since these attacks have various types due to the FL distributed nature, a good perception of attack characteristics and viable defense mechanisms is crucial. However, due to insufficient comprehensive research in this field, current defense methods are unsophisticated, and innovative defense approaches are still required[45].

Moreover, the incapability of FL methodology to access data on the client's side hinders detecting malicious update parameters sent by the malicious clients. Consequently, adversarial attacks form a significant setback facing FL. On the bright side, dangerous attacks on FL require white-box knowledge about the model (e.g., the aggregation method), which makes the scenario impractical. From the big picture (as shown in Fig 4), attacks could be divided into two classes (which will be discussed in the following):

1-Attacks on the FL model's security (the adversary's purpose is to modify the model's behavior)

2-Attacks on the FL model's privacy (that aim to infer critical information during the learning stage of the model).

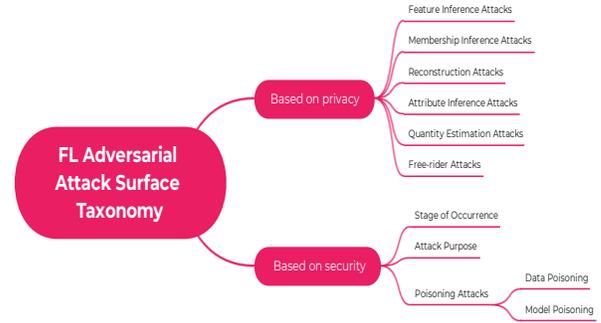

**Fig. 4.** Presentation of Federated Learning Adversarial Attack Surface taxonomy

### A. Taxonomy of Attacks on FL's Security

As mentioned, attacks on FL include a variety of forms; nevertheless, we can differentiate them based on their characteristics and the intruder's eventual aim.

#### 5) Stage of Occurrence

First, based on which stage of the learning process an attack occurs, we can classify attacks into two subcategories: training time and inference time. Training time attacks are the most popular and frequent type of these two. Training time attacks permit the adversary to accomplish arbitrary modification in training data in the data collection and data preparation phases[47]. Modifications in referred phases can lead to any modification in either behavior of the model or the outcome result of the model. Conversely, inference time attacks (Evasion Attacks) would be applied after the training phase. The primary purpose is misclassification and misprediction in the model.

#### 6) Attack Purpose

FL adversarial attacks also could be categorized as targeted or untargeted. Targeted attacks, known as Backdoor attacks, are highly severe due to their unexposed nature. One of the most applied attacks in this category is the Sybil attack. Genuinely, the model is exposed to multiple backdoor attacks in Sybil attacks, which are extremely hard to track[48]. In this case, multiple malicious client nodes inject unique patterns, or each node injects part of a specific pattern into the model. Consequently, given that Backdoor attacks were designed for secondary goals, there will be a broad variety of attacks based on backdoors that are beyond the purview of this study.

As opposed to Backdoor attacks that do not impact the performance, untargeted attacks only seek to cause damage to model performance. On top of that, untargeted attacks, such as free-riders or Byzantine attacks, do not steal data but rather produce random model updates and degrade performance by simply providing the server with a set of random update parameters learned on randomly updated data[49].

### B. Poisoning Attacks

Nearly all training time attacks use poisoned data on the client side to achieve their desire. Since these types of attacks include a broad spectrum of threats, we conducted more in-depth research on the kinds. The most significant criterion for distinguishing distinct poisoning attacks, the best matric, could be the ultimate objective of the attacker[48]. According to that, we can classify them as below subcategories:

● data poisoning

● model poisoning

Similarly, to their name, data poisoning attacks occur by maliciously poisoning training data. Az research indicates several approaches can be applied in this order. Label poisoning, sometimes called "label flipping," is one method. In contrast, sample poisoning may be accomplished by the addition of noise, a specific pattern to samples, or even some additional malevolent examples to mask the global model. It is essential to note that, according to[50], two elements may determine the success of an attack: the quantity of contaminated data and the number of attackers. Thus, the slope of success becomes steeper as the number of poisoned data increases.

As was stated previously, there is a second kind of poisoning attack known as model poisoning. Model poisoning involves the direct permutation of local update weights. Randomly generated weights or more complex methods like information leakage or optimization techniques might be used in such attacks[51].

Last but not least, we may develop another point of view regarding the representation of attack types according to their occurrence rates. Due to the extended duration of the training phase, attacks may be used either intermittently throughout a single round or continually for the whole training period.

### C. Taxonomy of Attacks on FL's Privacy

The majority of federated learning attacks encounter the model's security. However, some attacks intrude on FL privacy, called privacy or interference attacks. Although privacy attack frequency covers a small portion of attacks threatening FL, they are more vital and possess greater worth. They are intended to breach ML's participants and steal data[47]. They constitute a risk to the privacy of ML's training data and the privacy of individuals who consented to share their private information. The principle of federated learning comes with the idea that private data never leaves the source owner and only processes locally. As a result, traded models are liable to memorize the private training dataset. In the following, different taxonomy of privacy attacks will be explored.

#### 7) Feature Inference Attacks

These attacks attempt to retrieve the client's database in the FL task. The collected data type is mainly photos and plain text. Attacks in VFL gather private features and approach the most stringent setting in which the adversary has just the trained VFL model and updates parameters without requiring any topical knowledge. Although, in HFL attacks, the adversary targets information exchanged between participants.

#### 8) Membership Inference Attacks

A membership inference attack seeks to determine if a representative sample was employed during network training[52]. With this characteristic, they strive to learn about or recreate the training data for an ML system. Consequently, the individuals whose information was utilized for training the model may be at risk. The adversary conducting a membership inference attack may not be familiar with the model's training data or other internal parameters. Instead, the adversary is solely aware of the service employed to build the model or its algorithm and architecture (e.g., SVM, neural network).

Furthermore, an adversary can determine whether or not the precise data was trained using passive and active attacks[53]. In contrast to active attacks, passive ones do not interfere with the learning process and instead draw conclusions based on the observed changes to the model's parameters. However, as the FL model is being trained, active attackers can interfere with the training methodology and lead unsuspecting users to provide private information. Distributing malicious updates that trick the global model into revealing private information about other participants is one effortless method of doing this.

#### 9) Reconstruction Attack

In this category, the adversary attempts to recreate the samples using the parameters of the learning model; the authors of [54] presented the reconstruction attack in black box settings and worked throughout the training phase to reverse the original model. For the purpose of recovering inference data, [55] introduced the inversion attack model with the white-box and black-box perspectives. To distribute tasks, a malicious client could theoretically recover random inputs perfectly, even if it had no access to the data or calculations of the other clients.

#### 10) Attribute Inference Attacks

Attribute inference attacks, also known as property inference attacks, seek to misuse the FL model to infer the presence of an uncorrelated trait of any of the participants. Meaning it looks to infer a characteristic about a person or group of people that would not normally be expected to be shared; for instance, in an ML model used for face detection, the goal is to infer whether training photos contain faces with green eyes[56].

Moreover, by employing active/passive attacks, the adversary can gain specific attributes of the training data by acquiring more label information on the target traits. To make inferences, passive adversaries can monitor while the model is updated and afterward run a binary attribute classifier on the victim. Although the FL model's functionality is constrained by the attack condition of further training data, active attackers can trick the model into breaching data by training it to adequately discriminate between data and opt for only data with attributes.

#### 11) Quantity Estimation Attack

In this category, the attackers collect data from the models (that are weakly formed) to determine how many of each label existed. Label leaks in FL were described by [37]. This threat occurs when an inference is made based on the composition proportion of a dataset. The goal of the adversary is first to collect data from the intended subjects and then to successfully target a predetermined percentage of all consumers throughout all training tiers. The authors also started the attack in a passive mode, making it invisible to intrusion detection systems.

#### 12) Free-rider Attacks

Generally, any individual who benefits from any type of private or public resources without making any contribution[57] is assumed as a free-rider attacker. Free-rider attacks include faking involvement in the FL procedure to acquire the final aggregated model without contributing any data. This attack is crucial in federated learning applications when data is sparse, and the model has a significant commercial value.

## IV. DEFENSE STRATEGIES

Federated learning's distributed nature makes almost every defense in typical-centric machine learning methods insufficient. Traditional defenses, such as homomorphic encryption (HE) and secure multi-party computation (SMC), do not have enough scalability to be adopted on FL. Although differential privacy (DP) has enough scalability, it has specific flaws that make it not a viable and robust solution to FL attacks. Therefore, as demonstrated in Fig 5, we demonstrated all present-day defenses in this section[58,59].

Furthermore, the proposed defenses have three possible focuses: the server, the client, and the communication channel (specific categories may reveal a combination of characteristics). Diverse research points to the fact that most studies have advocated server-side techniques because the server would be the most acceptable element under a federated paradigm. Even though it is plausible that communication channels will be accessible, it is challenging to offer a scalable strategy when dealing with such a considerable number of clients[60].

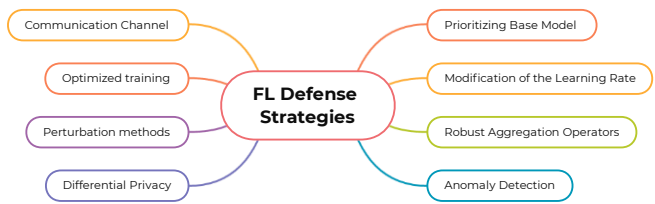

**Fig. 5.** Illustration of Federated Learning Existing Defense Methodologies

### A. Prioritizing Base Model

A very primitive approach in defense mechanisms (applies to the server) would be this approach. It illustrates a model that already has a good knowledge of data and does not require many updates and changes[60]. Consequently, the base model always weighs far more than the clients' data influencing the model, making the assumption impractical.

### B. Modification of the Learning Rate

The server has the upper hand since it is responsible for setting the learning rate, which regulates the relative importance of the global model's previous version and the aggregate of client model updates. As a result, the server controls who may take part in which aspects of the model updating process. The literature has used this decomposition method to protect the federated model against adversarial attacks[61].

PDGAN[61] is a prime example of this defense methodology; it is an innovative strategy for dealing with poisoning attacks. Using the produced data, PDGAN is able to correctly rebuild the samples from updates' parameters and probe the client model accuracy. So, if a trainee's accuracy goes below a certain level, they are flagged as a potential threat and dropped out of the process.

### C. Robust Aggregation Operators

This approach mitigates the poisoning attacks by substituting estimators instead of the mean on the outliers or extreme values. Median[62], Norm thresholding, Adaptive Federated Averaging[21], and Game-theory[63] are some illustrations of aggregation operators that are based on more reliable estimators. However, the degree to which the parameters of each client update should affect the global model during training is something that must be decided separately for each approach. In some cases, the substitution is just mathematical; in others, creativity is required. However, innovation in this stage could cause disguise. For instance, results show that the model drops malicious or weak client data in the Adaptive Federated Averaging solution, which employs a Markov Model to calculate weights for client data.

### D. Anomaly Detection

These defenses recognize hostile clients as aberrant data in local model updates and exclude them from the aggregate. This approach uses modified univariate or multivariate anomaly detection ML algorithms[64,65]. The primary issue with techniques of this kind is that predictions are often high-dimensional and derived from neural networks.

This category of defenses uses analytical and statistical techniques to discover occurrences that deviate from predicted patterns or behaviors. Anomaly detection algorithms might be employed to discover misbehaving clients in FL environments. The FL server evaluates individual updates and their impact on the global shared model in order to detect attacks such as poisoning attacks. However, when it comes to targeted backdoor attacks, these systems are quite vulnerable[66].

### E. Differential Privacy

Adding noise parameters into system variables may be one of the preliminary ideas to increase the security of machine learning techniques. Nevertheless, it remains one of the most robust and feasible approaches for mitigating FL adversaries' attacks on both the server and client sides[67]. Typically, DP defenses are developed to confront server-side privacy attacks; they also prevent malicious attempts against clients. Differential privacy has been adopted successfully in various ML approaches (e.g., SVM, LRR); hence, taking DP into Federated learning is quite challenging.

On the downside, in the case of imbalanced data, differential privacy bitterly diminishes FL model performance [68,69]. To address this issue, deploying DP to the aggregate operator might greatly assist. For instance, [70] introduces CentralDP, a differentially private aggregation operator that eliminates outliers by clipping the norm of the prediction update parameters (much like the Norm thresholding operator) and adds Gaussian noise calibrated to the clip. It has some conceptual similarities to the resilient aggregation operators and was originally derived from the FedAvg operator.

Differential privacy, as stated at the beginning of this section, is one of the few ways to improve the federated learning paradigm. That is due to the fact that not only it is a viable solution, but also it can be implemented on both the server and client sides. Although DP is designed to combat privacy attacks on the server, research indicates DP can lessen adversarial attacks on the client side. Accordingly, innovative paradigms such as Local DP[71] and f-DP[72] were introduced to advance this subject.

## F. Perturbation Methods

In this model (applies to the client), the most vulnerable component would be subjected to a noise function with the objective of reducing the amount of information an adversary may retrieve. The primary focus of works derived from this technique is on customers' local datasets or shared update parameters. One of the most impressive case studies for perturbation methods would be [70]. The authors suggest a novel perturbation function applied to client data, with the ability to detach the perturbation being limited to the server. This method eliminates the possibility of data recovery and renders data interception ineffective.

## G. Optimized training

Optimizing benign client training is another solution to defeat adversarial attacks. According to[73], fine-tuning can be performed to maximize harmless customers' aggregate influence. The authors utilized specific metrics from clients' networks to determine safe participants. In this way, they neutralized backdoor attacks at the cost of performance degradation.

Another common approach is Adversarial training which defends the federated model from outside threats. In contrast to previous defensive tactics, adversarial training tries to enhance the internal robustness of federated models. According to[74], a learning model can "pivot" on the sensitive characteristics in order to create predictions that are independent of those features in the training data.

However, adversarial training has a long way to go before it can handle adversarial attacks flawlessly. Existing adversarial training methods can provide a model with an MNIST worst-case accuracy of approximately 90%. In adversarial training, ML engineers retrain their models using adversarial examples to make them robust against data perturbations[75]. Each training instance must be analyzed for an adversarial flaw, and then the model should be trained again. Hence, adversarial training is a costly and time-consuming technique. Therefore, optimizing the detection and correction of adversarial flaws in machine learning models is a potential focus for researchers[76].

## H. Communication Channel

All safe FL implementations fall under the category of communication channel defenses. They allow groups to carry out a global objective, even in exposure to potentially harmful intruders[77]. The challenge is that although the inputs to global computing activities are protected, outputs are disclosed to at least some parties and are not provided with a safety guarantee. As a result, the confidentiality of outputs may be compromised. Experimental works against privacy attacks led to this strategy, which is based on the idea that if an attacker cannot access specific output parameters, the attack may fail. In this direction, the concept of restricting access to information comes as a pivotal core to communication channel defenses. State-of-the-art developed protocols in this group of defenses are named SMPC and BFL [78,79].

SMPC is the abbreviation for Secure Multi-party Computation protocol. It is founded based on Homomorphic Encryption providing secret sharing[80]. SMPC could be embodied on partial/final updates or during aggregation. Given that, due to the HE basis of SMPC, all encryptions would be settled with a single key methodology. conceptually, this may lead to the risk of compromised key and a single point of failure.

However, unlike SFL protocols, BFL or Blockchain-based FL permits a highly scalable and decentralized FL environment without a single point of failure issue[81]. Nevertheless, we might contemplate that although this approach might seem to function, we are still not immune to the dangers lurking in the blockchain, such as forking attacks and reentrancy attacks.

## V. Challenges of Resolving Risks posed by federated learning

As FL is in its prime stages of development, it has several unresolved issues that demand careful consideration. Due to the dissimilarity between the federated setting and traditional difficulties, the challenges pose an intriguing research paradigm to the researchers. This part addresses some of the most significant obstacles facing federated learning [13,82].

## A. Security and Privacy Threats

As opposed to learning from centralized data in data centers, FL apps frequently prioritize privacy[2]. FL aids in the protection of user data by sharing only updated models (including gradient information) as opposed to the complete dataset. However, training local models and then transferring changes to a centralized server or a third party might potentially compromise private information[83].

While efforts have been made recently to enhance the privacy of federated learning using approaches such as DP and safe multi-party computing, these solutions might sacrifice system performance or model accuracy for privacy. Identifying and evaluating trade-offs on a theoretical and practical level is a crucial step toward developing private federated learning systems[84].

## B. The Trade-Off in Defenses

For each proposed defense technique, there is at least one weakness that could be transformed into an adversarial attack[70]. Therefore, it is not easy to devise a defense against either of these problems:

- Any inherent vulnerabilities in an FL situation that an attacker could exploit.
- A defense capable of defending against any attack.

In most defenses, we find it challenging to keep a balance between preventing attacks and slowing performance on the primary objective. In models based on DP, for instance, a considerable amount of noise must be included to assure data privacy, which dramatically diminishes the model's performance[85].

Therefore, developing more efficient DP methods and extending them to defend against all adversarial attacks would be pretty challenging. This issue also arises with client-filtering-based defenses when abundant clients have filtered away, resulting in information loss during the aggregation procedure. In short, finding a balance between preventing an attack and losing or tainting customer information is the primary obstacle in front of FL[86].

## C. Generalized FL

It has been shown that HFL is the target of the most adversarial attacks against the federated structure. While research on privacy attacks in VFL does exist, more work remains to be done in identifying and understanding the vulnerabilities in other rapidly growing categories of FL, such as VFL or FTL, in terms of information leakage of attacker possibilities. For this reason, we anticipate that detecting VFL and FTL attacks and investigating defenses will take center stage in the future[87].

### D. Non-Independent and Identically Distributed (Non-IID) Assumptions

Due to the non-independent and identically distributed nature of the training data on clients, it is nearly impossible to distinguish adversarial clients from distinct but valuable participants[72,88]. Therefore, using anomaly detection algorithms appropriate for non-IID data or methodologies not discriminating between data distributions is a frequent strategy. However, there are still difficulties in distinguishing between clients with a highly skewed distribution and adversarial clients in most circumstances.

### E. Heterogeneity Challenge

Network connectivity, processing capacity, source power, communication, hardware capacity, and so on all vary from node to node in federated networks, making them inherently heterogeneous. Consequently, integrating data from multiple heterogeneous devices in FL may be a severe challenge.

Moreover, trust problems and inefficiency are becoming more apparent as the local workforce becomes more diverse and complicated. So evidently, research points out that federated learning compared to machine learning, has a more prolonged convergence time. That being so, efforts comparable to [89] are necessary to lower the convergence time by emulating hierarchical and centralized approaches.

### F. Communication Challenge

High latencies and limited data transfer rates are critical challenges to the FL paradigm (where minimal latency is required for rapid learning using the backpropagation technique). The learning process is significantly slowed down when countless devices are used to train the algorithm. This challenge is exacerbated when bandwidth is considered a technological concern. Since most places are equipped with Wi-Fi or 4G that cannot handle FL's data demands, significant delays throughout the learning process will be caused. In other words, the communication bottleneck results from the increased processing capability of FL devices without corresponding increases in bandwidth. Therefore, 5G and B5G have to be employed in the FL, and also models should be compressed and quantized to deal with communication costs.

## VI. Discussion

Federated Learning is a distributed machine learning approach that trains models without centralizing data. This paper presents a complete overview of the security and privacy challenges associated with FL, and highlights the importance of addressing these challenges to increase trust in FL and enable broader adoption. This study also outlines the attack surface of FL and defense measures to evaluate its robustness and reliability.

The paper's examination of the frequency of security and privacy issues in FL is one key contribution of this survey. Communication bottlenecks, poisoning, and backdoor attacks are identified as the three most major security dangers, while privacy concerns are found to be less prevalent. In addition, this article offers multiple avenues for future research to improve the security and privacy of FL in real-world scenarios. To be more specific, this survey advises investigating adversarial training as a way of defense against malicious clients and implementing more severe techniques to guard against attackers in federated learning. Further, it calls for additional research on privacy-preserving aggregating approaches and the influence of FL on fairness and bias in machine learning.

All in all, by providing a comprehensive and instructive review of the existing security and privacy aspects of FL, we firmly believe that FL's ultimate goal targets gathering and learning a broad range of data. Nonetheless, given FL's increasing adoption, more security specialists and machine learning experts should discuss state-of-the-art methodologies to enhance FL's security status. Considering existing research in FL security and privacy, combining methods is recommended to bring more satisfying results.

## VII. Conclusion and Future Research

Federated Learning arises as an innovative machine learning solution to increase computing and privacy requirements. This approach enables numerous nodes to construct a global model collaboratively to address crucial issues such as security and privacy, access rights, and heterogeneous data access, among others. Recommendation systems, the IoT, banking and finance, NLP, and healthcare are some fields that utilize FL. However, lack of consideration towards its security, privacy, and open challenges resulted in FL's moderated development and not reaching its full potential.

That said, this new paradigm has presented additional obstacles, especially regarding adversarial attacks. Whereas FL claims to preserve data security and privacy, it provides adversaries with data breaches and misclassification. Accordingly, FL has allowed opponents to misuse it, utilizing poisoned data, manipulated model updates, data leaks, and model compromises.

This study highlights the vulnerable surface of the federated learning process. The taxonomy illustrates the attack surface of FL within the perspective of security and privacy targeted attacks. During that, feasible attacks' categories and impacts are also described. Thereon, contemporary defensive techniques are outlined to address privacy and security risks, along with their weaknesses and directions for improvement. Next, we contended that the diversity of attacks surpassed the existing FL defensive mechanisms.

To summarize, this survey first described a comprehensive study of the adversarial FL environment. Secondly, a classification for privacy and security flaws is presented. Additionally, cutting-edge defensive strategies concerning FL attacks and well-known models against security and privacy threats are explored. Finally, we identify prospects for strengthening protection systems and critical insights into the FL security concerns.

In conclusion, this research suggests that additional research is required to protect FL against adversarial attacks since most research is concentrated in a few key areas, such as healthcare and the Internet of Things. However, this learning paradigm might be used in a variety of fields, including food delivery, virtual reality applications, public safety, accident reporting, traffic control, and monitoring. We reckon data security and privacy must come first for all other methods to matter. Therefore, future studies will take a more holistic approach to this problem, offering more tangible, laudable traits.